\documentclass[prl,amsfonts,amssymb,amsmath,showpacs,twocolumn]{revtex4}
\date{\today}
\begin{document}

\title{Gravity-induced vacuum dominance}
\author{William C.\ C.\ Lima}\email{william@ursa.ifsc.usp.br}
\affiliation{Instituto de F\'\i sica de S\~ao Carlos,
Universidade de S\~ao Paulo, Caixa Postal 369, CEP 15980-900, 
S\~ao Carlos, SP, Brazil}
\author{Daniel A.\ T.\ Vanzella}\email{vanzella@ifsc.usp.br}
\affiliation{Instituto de F\'\i sica de S\~ao Carlos,
Universidade de S\~ao Paulo, Caixa Postal 369, CEP 15980-900, 
S\~ao Carlos, SP, Brazil}

\begin{abstract}
It has been widely believed that, except in very extreme situations, the
influence of gravity on quantum fields should amount to just small,
sub-dominant contributions. This view seemed to be
endorsed by the seminal results obtained over the last decades in the context 
of renormalization of quantum fields in curved spacetimes. Here, however, 
we argue that this belief is {\it false} by showing that there exist 
{\it well-behaved} spacetime evolutions where the vacuum energy density 
of {\it free} quantum fields is {\it forced}, by the very same background
spacetime, to become dominant over {\it any} classical energy-density 
component. This semiclassical gravity effect finds its roots in the 
{\it infrared} behavior of fields on curved spacetimes. By estimating the 
time scale for the vacuum energy density to become dominant, and therefore 
for backreaction on the background spacetime to become important, we argue 
that this vacuum dominance may bear unexpected astrophysical and cosmological 
implications.
\end{abstract}


\maketitle

In the absence of a full quantum gravity theory, the influence of gravity on 
quantum fields can be properly analyzed only in the semiclassical approximation, 
in which matter (and other interaction) fields are quantized on classical 
background spacetimes. This semiclassical approach, known as {\it quantum field 
theory in curved spacetimes} 
(QFTCS)~\cite{Parker68,Parker69,BD82,Fulling89,Wald94,Parker09}, gives 
meaningful results as long as it deals with situations far away from the Planck 
scale. The {\it Hawking effect}~\cite{Hawking74,Hawking75}, according to which 
black holes should emit a thermal bath of particles, 
provides an example of the strength of the QFTCS formalism. 
However, in spite of its conceptual importance, it has been widely believed 
that except in very extreme situations (near singularities, Cauchy horizons, 
tiny black holes), the influence of gravity on quantum phenomena should amount 
only to small, sub-dominant contributions. Here we argue that this commonly 
held belief is {\it false}. For the sake of simplicity, we focus on the vacuum 
energy density of a {\it free} quantum scalar field and show that on some
{\it well-behaved} spacetimes it can become {\it dominant} over {\it any}
classical energy-density component, even though it is bound to remain finite 
everywhere. We also show, by performing a simple estimate, that the 
{\it natural} time scale for this semiclassical gravity effect to become 
important, if it is triggered, is of tiny fractions of a second in some 
astrophysical contexts, while in cosmological contexts it would be of a few 
billion years.

Let us begin by considering a real, free scalar field $\Phi$ with mass $m$ 
satisfying the usual Klein-Gordon equation with the additional coupling to 
the scalar curvature $R$:
\begin{eqnarray}
\left(-\square +m^2 +\xi R
\right) \Phi=0,
\label{FE}
\end{eqnarray}
where $\xi$ is a real constant. (We adopt natural units in which $\hbar=c=1$, 
unless stated otherwise.) The associated quantum field $\hat{\Phi}$ is 
formally written as
$\hat{\Phi}=\int d\mu(\alpha) 
[
\hat{a}_\alpha u^{(+)}_\alpha +\hat{a}^{\dagger}_\alpha u^{(-)}_\alpha
]$,
where $u^{(+)}_\alpha$ and $u^{(-)}_\alpha\equiv(u^{(+)}_\alpha)^\ast$ are 
positive- and negative-norm solutions of Eq.~(\ref{FE}), respectively, which 
together form a {\it complete} set of normal modes, satisfying
$(
u^{(+)}_{\alpha},u^{(+)}_{\beta} 
)_{\rm KG}=-(
u^{(-)}_{\alpha},u^{(-)}_{\beta})_{\rm KG}
=\delta(\alpha, \beta)
$ and
$
(
u^{(+)}_{\alpha},u^{(-)}_{\beta} 
)_{\rm KG}=0
$,
with $\delta(\alpha, \beta)$ being the Dirac's ``delta function''
associated with the measure $\mu(\alpha)$ on the set of ``quantum numbers'' 
$\alpha$. Recall that the Klein-Gordon inner product defined on the space 
${\cal S}$ of complex solutions of Eq.~(\ref{FE}) is given by
$(
u,v 
)_{\rm KG}:=
i \int_{\Sigma}d\Sigma \;n^a\left[
u^\ast \nabla_a v-v \nabla_a u^\ast
\right]$,
where $d\Sigma$ is the proper-volume element on the Cauchy surface $\Sigma$
and $n^a$ is the future-pointing unit vector field orthogonal to $\Sigma$.
The operators $\hat{a}_\alpha$ and $\hat{a}^\dagger_\alpha$ are taken to
satisfy the canonical commutation relations (CCR) 
$[\hat{a}_\alpha,\hat{a}^\dagger_{\beta}
]=\delta(\alpha,\beta)$,
$[\hat{a}_\alpha,\hat{a}_{\beta}
]=0$,
from where the mode-annihilation and -creation interpretation follows, as 
well as the Fock-space construction based on the ``vacuum'' state $\left| 
0 \right>$ defined through $\hat{a}_\alpha \left| 0 \right>=0$ for all 
$\alpha$. Obviously, the choice of the solutions to constitute the 
positive-norm modes $u^{(+)}_\alpha$ is far from unique, and
different choices can lead to different (i.e., unitarily {\it inequivalent}) 
Fock spaces of states where the CCR is implemented. In the absence of a 
time-like symmetry, with respect to which a preferred notion of
positive-frequency solutions can be defined, there is no natural way of 
picking one space out of the infinite possibilities. As a consequence, no 
natural notion of {\it particles} exists in a general curved spacetime. 
This, however, poses no impediment to the formalism of QFTCS, as is well 
known.

The effect we shall discuss here does {\it not} rely on this 
``indeterminacy'' of the particle concept. Therefore, in order to avoid 
unnecessary complications we shall assume a globally hyperbolic spacetime 
which is conformally static in both the asymptotic past and future. To be 
even more conservative, we focus attention on a spacetime which is 
conformally {\it flat} in the asymptotic past:
\begin{eqnarray}
ds^2\sim
\left\{
\begin{tabular}{ll}
$f_{in}^2(-dt^2+d\vec{x}^2)$ &  , asymp.\ past\\
$f_{out}^2 (-dt^2+ h_{i j} dx^i dx^j)$ &  , asymp.\ future
\end{tabular}
\right. 
\label{ds2}
\end{eqnarray}
where $f_J=f_J(t, \vec{x})>0$, $J\in \{in,out\}$,
are smooth functions and $h_{ij}=h_{ij}(\vec{x})$, $i,j=1,2,3$, are the 
components of an arbitrary spatial metric. (We use the same labels $t$ 
and $\vec{x}=(x^1,x^2,x^3)$ for coordinates in the asymptotic past and 
future only for simplicity; they are obviously defined on non-intersecting 
regions of the spacetime.) In each of these asymptotic regions the field 
$\Phi$ can be written as $\Phi=\tilde{\Phi}/f_J$, where 
$\tilde{\Phi}$ satisfies
\begin{eqnarray}
-\frac{\partial^2}{\partial t^2}\tilde{\Phi}= -\Delta_J\tilde{\Phi}+
V_J\tilde{\Phi},
\label{tildeKG}
\end{eqnarray}
where $\Delta_{in}$ is the usual (flat) Laplace operator, $\Delta_{out}$ 
is the Laplace operator associated with the spatial metric $h_{ij}$, and 
the effective potential $V_J$ is given by
\begin{eqnarray}
V_J&=&\frac{(\Delta_{J} f_{J}-\ddot{f}_{J})}{f_{J}}+f_J^2(m^2+\xi R)
\nonumber \\
&=&(1-6 \xi)\frac{(\Delta_{J} f_{J}-\ddot{f}_{J})}{f_{J}}
+f_{J}^2m^2+\xi K_J,
\label{Veff}
\end{eqnarray}
with $K_{in}=0$, $K_{out}=K_{out}(\vec{x})$ the scalar curvature associated 
with the spatial metric $h_{ij}$, and the dots denoting differentiation with 
respect to the variable $t$.

Although Eq.~(\ref{tildeKG}) is already in a form upon which our main line 
of reasoning could be constructed, let us simplify our analysis further by 
assuming that  $V_{in}=0$ and $V_{out}$ does not depend on $t$, 
$V_{out}=V_{out}(\vec{x})$. This is certainly not the case in general for 
spacetimes whose metric satisfies Eq.~(\ref{ds2}), but there are very 
interesting situations which do satisfy this condition: (i) the massless 
($m=0$) field with arbitrary coupling $\xi$ in spacetimes which are 
asymptotically flat in the past and asymptotically static in the future
[$f_{in}=1$ and $f_{out}=f_{out}(\vec{x})$], as those describing the 
formation of a static star from matter initially scattered throughout space, 
and (ii) the massless, conformally coupled field ($m=0$ and $\xi=1/6$). 
With this assumption for the potential, two different sets of positive-norm 
modes, $u^{(+)}_{\vec{k}}$ and $v^{(+)}_\alpha$, can be naturally defined by 
the requirement that they are the solutions of Eq.~(\ref{FE}) which satisfy 
the asymptotic conditions:
\begin{eqnarray}
u^{(+)}_{\vec{k}}
\;\sim^{^{\!\!\!\!\!\!\!\! \rm past}}
(16\pi^3 \omega_{\vec{k}})^{-1/2}\;f_{in}^{-1}\;
e^{-i(\omega_{\vec{k}} t-\vec{k}\cdot \vec{x})}
\label{uplus}
\end{eqnarray}
and
\begin{eqnarray}
v^{(+)}_{\alpha}
\;\;\sim^{^{\!\!\!\!\!\!\!\!\!\! \rm future}}
(2\varpi_{\alpha})^{-1/2}\;f_{out}^{-1}\;
e^{-i\varpi_{\alpha} t} F_{\alpha}(\vec{x}),
\label{vplus}
\end{eqnarray}
where $\vec{k}\in {\mathbb R}^3$, $\omega_{\vec{k}}:=\|\vec{k}\|$, 
$\varpi_\alpha>0$, and $F_\alpha(\vec{x})$ are solutions of
\begin{eqnarray}
[-\Delta_{out}+V_{out}(\vec{x})]F_{\alpha}(\vec{x})=\varpi_\alpha^2 
F_\alpha(\vec{x})
\label{Falpha}
\end{eqnarray}
satisfying the normalization
\begin{eqnarray}
\int_{\Sigma_{out}} d^3x \sqrt{h} \;F_\alpha(\vec{x})^\ast 
F_\beta(\vec{x})=\delta(\alpha,\beta)
\label{norm}
\end{eqnarray}
on a Cauchy surface $\Sigma_{out}$ in the asymptotic future. (Each 
$F_\alpha$ can be chosen to be real with no loss of generality.)

The fact that in general the modes $v^{(+)}_\alpha$ {\it cannot} be 
expanded in terms of $u^{(+)}_{\vec{k}}$ alone ($u^{(-)}_{\vec{k}}$ 
might be needed) is responsible for the almost-forty-year-old effect 
of particle creation due to the (change in the) gravitational background: 
the vacuum state $\left|0\right>_{in}$ associated with the modes 
$u^{(+)}_{\vec{k}}$, which represents absence of particles in the 
asymptotic past, represents a particle-filled state according to the 
natural notion of particles in the asymptotic future (associated with 
$v^{(+)}_\alpha$).  This stands at the root of the Hawking effect and of 
particle creation in expanding universes~\cite{Parker68,Parker69}. Here, 
however, we want to call attention to a different effect, independent of 
particle creation, which seems to have passed unnoticed in the general 
context: there are {\it reasonable} situations where the modes 
$v^{(+)}_\alpha$, given in Eq.~(\ref{vplus}), together with 
$v^{(-)}_\alpha$ {\it fail} to form a {\it complete} set of normal modes. 
This happens whenever the operator $[-\Delta_{out}+V_{out}(\vec{x})]$ in 
Eq.~(\ref{Falpha}) happens to possess normalizable [i.e., satisfying 
Eq.~(\ref{norm})] eigenfunctions with {\it negative} eigenvalues, 
$\varpi_\alpha^2=-\Omega_\alpha^2<0$. In this case, additional 
positive-norm modes $w^{(+)}_\alpha$ with the asymptotic behavior
\begin{eqnarray}
w^{(+)}_\alpha
\;\;\sim^{^{\!\!\!\!\!\!\!\!\!\! \rm future}}
\frac{\left(e^{\Omega_{\alpha} t-i\pi/12}+e^{-\Omega_{\alpha} t+i\pi/12}
\right) 
F_{\alpha}(\vec{x})}{\sqrt{2\Omega_{\alpha}}\;f_{out}(t,\vec{x})}
\label{wplus}
\end{eqnarray}
and their complex conjugates $w^{(-)}_\alpha$ are {\it necessary} in order 
to expand an arbitrary solution of Eq.~(\ref{FE}). As a direct consequence, 
at least some of the in-modes $u^{(\pm)}_{\vec{k}}$ (typically those with 
low $\omega_{\vec{k}}$) eventually undergo an exponential growth (assuming 
that $f_{out}$ remains polynomially bounded). This {\it asymptotic 
divergence} is reflected on the unbounded increase of the vacuum 
fluctuations, 
\begin{eqnarray}
\left<\Phi^2\right>
\;\;\sim^{^{\!\!\!\!\!\!\!\!\!\! \rm future}}
\frac{\kappa\;e^{2\bar\Omega t}}{2 \bar\Omega}
\left(\frac{\bar{F}(\vec{x})}{f_{out}(t,\vec{x}) }\right)^2
\left[
1+
{\cal O}(e^{-\epsilon t})\right],
\label{Phi2}
\end{eqnarray}
where $\bar{F}(\vec{x})$ is the eigenfunction of Eq.~(\ref{Falpha}) 
associated with the lowest {\it negative} eigenvalue allowed,  
$\varpi_\alpha^2=-\bar\Omega^2$,  $\epsilon$ is some positive constant, and
$\kappa$ is a dimensionless constant (typically of order unity) whose exact 
value depends {\it globally} on the spacetime structure (since it crucially 
depends on the projection of each $u^{(\pm)}_{\vec{k}}$ on the mode
${w}_{\alpha}^{(\pm)}$ whose $\varpi_\alpha^2=-\bar\Omega^2$; $\kappa$ also 
depends on the initial state, here assumed to be the vacuum 
$\left| 0 \right>_{in}$).

As one would expect, these wild quantum fluctuations give an important 
contribution to the vacuum energy stored in the field. In fact, the 
expectation value of its energy-momentum tensor, $\left<T_{\mu\nu}\right>$, 
in the asymptotic future is found to be dominated by this exponential growth:
\begin{eqnarray}
\left<T_{00}\right>
&\sim^{^{\!\!\!\!\!\!\!\!\!\! \rm future}}&\!\!\!\!
\left< \Phi^2\right>\left\{
\frac{(1-4 \xi)}{2}\left(
\bar\Omega ^2+\frac{(D\bar{F})^2}{\bar{F}^2}
+
m^2 f^2 +\xi K \right)\right. \nonumber \\
& &
+(1-6 \xi)
\left(
\frac{2 \xi \ddot{f}}{f}- \frac{2\xi D^2 f}{f}
+\frac{\dot{f}^2}{2f^2}
-\frac{\bar\Omega\dot{f}}{f}\right.\nonumber \\
& &\left. \left.
+\frac{(D f)^2}{2f^2}
-\frac{D_if D^i\bar{F}}{f \bar{F}}
\right)
+{\cal O}(e^{-\epsilon t})
\right\},
\label{T00}
\end{eqnarray}
\begin{eqnarray}
\left<T_{0i}\right>
&\sim^{^{\!\!\!\!\!\!\!\!\!\! \rm future}}&\!\!\!\!
\left< \Phi^2\right>\left\{
(1-4 \xi)
\frac{\bar\Omega D_i\bar{F}}{\bar{F}}
+(1-6 \xi)
\left(
\frac{\dot{f}D_if}{f^2}
\right.
\right. \nonumber \\
& &
\left. \left.
- \frac{\dot{f}D_i\bar{F}}{f\bar{F}}
-\frac{\bar\Omega D_if}{f}\right)
+{\cal O}(e^{-\epsilon t})
\right\},
\label{T0i}
\end{eqnarray}
\begin{eqnarray}
\left<T_{ij}\right>
&\sim^{^{\!\!\!\!\!\!\!\!\!\! \rm future}}&\!\!\!\!
\left< \Phi^2\right>\left\{
(1-2 \xi)\frac{D_i\bar{F} D_j\bar{F}}{\bar{F}^2}
-2 \xi \frac{D_i D_j \bar{F}}{F}
+\xi \widetilde{R}_{ij}\right.
\nonumber \\
& &+
\frac{(1-4\xi)h_{ij}}{2}\left(
\bar\Omega ^2-\frac{(D\bar{F})^2}{\bar{F}^2}
-
m^2 f^2 -\xi K
\right)
\nonumber \\
& &
+(1-6 \xi)
\left[\frac{D_ifD_jf}{f^2}
-\frac{D_if D_j\bar{F}}{f\bar{F}}-\frac{D_jf D_i\bar{F}}{f\bar{F}}
\right.
\nonumber \\
& &
+h_{ij}
\left(
\frac{2\xi D^2 f}{f}-\frac{2 \xi \ddot{f}}{f}
+\frac{\dot{f}^2}{2f^2}
-\frac{\bar\Omega\dot{f}}{f}
-\frac{(D f)^2}{2f^2}
\right.\nonumber \\
& &\left. \left.\left.
+\frac{D_kf D^k\bar{F}}{f \bar{F}}
\right)\right]
+{\cal O}(e^{-\epsilon t})
\right\},
\label{Tij}
\end{eqnarray}
where $D_i$ is the derivative operator compatible with the metric 
$h_{ij}$ (so that $\Delta_{out}=D^2$), $\widetilde{R}_{ij}$ is the 
associated Ricci tensor (so that $K_{out}=h^{ij}\widetilde{R}_{ij}$), 
and we have omitted the subscript $out$ in $f_{out}$ and $K_{out}$ for
simplicity. The Eqs.~(\ref{T00}-\ref{Tij}), together with 
Eq.~(\ref{Phi2}), imply that on time scales determined by 
$\bar\Omega^{-1}$, the vacuum fluctuations of the field should  
overcome any other classical source of energy, therefore taking control 
over the evolution of the background geometry through the semiclassical 
Einstein equations (in which $\left<T_{\mu \nu}\right>$ is included as a 
source term for the Einstein tensor). We are then confronted with a 
startling situation where the quantum fluctuations of a field, whose 
energy is usually negligible in comparison with classical energy 
components, are {\it forced} by the background spacetime to play a 
{\it dominant} role. 

We are still left with the task of showing that there exist indeed 
well-behaved background spacetimes in which the operator 
$[-\Delta_{out}+V_{out}(\vec{x})]$ possesses negative eigenvalues
$\varpi_\alpha^2<0$, condition on which depends all the discussion 
presented above. Experience from usual quantum mechanics tells us that 
this typically occurs when $V_{out}$ gets sufficiently negative over a 
sufficiently large region. It is easy to see from Eq.~(\ref{Veff}) that, 
except for very {\it special} geometries (as the flat one), one can 
generally find appropriate values of $\xi\in {\mathbb R}$ which make 
$V_{out}$ as negative as would be necessary in order to guarantee the 
existence of negative eigenvalues. Therefore, the question is not 
{\it if} negative eigenvalues are possible, but {\it how natural} are 
the scenarios in which they appear. For massless fields with coupling 
$\xi$ of order unity, $V_{out}$ is of order $R$ [see Eq.~(\ref{Veff})], 
which in turn is of order $8 \pi G\rho_c$ (assuming the validity of the 
{\it classical} Einstein equations), where $\rho_c$ is the energy density 
of the classical matter governing the spacetime evolution and $G$ is 
Newton's constant. Note also that we can manipulate the sign of $V_{out}$ 
by choosing $\xi$ properly (but still with values of order 1). Combining 
all these observations suggests that background geometries 
associated with matter distributions whose density variations are of order 
$\delta \rho_c$ over regions of typical linear size $L$, satisfying 
$8 \pi G \delta \rho_c L^2 \sim 1$ or larger, are promising candidates where
a massless field with appropriate coupling $\xi$ (with $|\xi|\sim 1$) would 
exhibit  the vacuum-dominance effect presented above. Recovering units
appropriate in different contexts, we 
have
\begin{eqnarray}
\frac{8\pi G\delta\rho_c L^2}{c^2}&\approx & \left(\frac{\delta\rho_c}
{10^{15}~{\rm g}/{\rm cm}^3}\right)
\left(\frac{L}{7~{\rm km}}\right)^2\nonumber \\
&\approx &
\left(\frac{\delta\rho_c}{\rho_{m0}}\right)
\left(\frac{L}{4.7\times 10^3~{\rm Mpc}}\right)^2 \sim 1,
\label{cond}
\end{eqnarray}
where $\rho_{m0}\approx 2.5 \times 10^{-30}~{\rm g}/{\rm cm}^3$ is the matter 
density (baryonic and dark) averaged over the observable universe, whose 
linear size is comparable to the Hubble length $4.1\times 
10^3$~Mpc~\cite{PDG}. 

This crude estimate serves only to {\it suggest} the 
scenarios in which the vacuum-dominance effect might play some role: 
{\it compact objects}~\cite{Lattimer07} 
and {\it cosmology}. 
Only a thorough analysis can properly reveal the 
relevance of the mechanism in each of these contexts. Notwithstanding,
although the main goal of this letter is to lay the general basis of the 
mechanism, next we 
summarize the results of a detailed analysis
performed in the simplest (non-trivial) instance where the vacuum dominance 
is found to be
triggered: the background geometry of
a uniform-density, spherically-symmetric compact 
object~\cite{LMV}. In such an idealized case, the Tolman-Oppenheimer-Volkoff 
equation
(which relates pressure and density inside the object) can be analytically 
solved (see, e.g., Ref.~\cite{Wald84}), from where the background geometry
[$f_{out}$ and $h_{ij}$ in Eq.~(\ref{ds2})]
can be calculated and substituted into the expression for $V_{out}$, 
Eq.~(\ref{Veff}). Then, it is simply a matter of verifying (numerically) 
the existence of bound eigenfunctions for the operator 
$(-\Delta_{out}+V_{out})$ appearing in Eq.~(\ref{Falpha}). After 
performing this procedure for several values of the compact-object mass $M$ 
and
radius $r_o$, it is found that there always exist {\it classically-stable} 
compact-object configurations (i.e., with $M/r_o<4/9$ in geometric units)
which awake the vacuum energy of massless fields with {\it any} value of 
$\xi>1/6$ or $\xi < \xi_0$ (with $\xi_0 \approx -2$). Preliminary 
results~\cite{LMV} show that more realistic compact objects (like some neutron 
stars) can also trigger the effect for massless fields with appropriate 
couplings. This leads to an interesting (and rare) possible
interconnection between 
observational astrophysics and semiclassical gravity, where
the observation of stable neutron-star configurations may rule out 
the existence of certain fields in Nature. 

Back to the general context, the time scale $\bar\Omega^{-1}$ (typically or 
order 
$|V_{out}|^{-1/2}$), which determines how sharp would be the transition from 
classical to vacuum dominance, can be estimated as being given by $L$ when 
condition~(\ref{cond}) is verified. Therefore, for compact objects we have
$\bar\Omega^{-1}\sim 10^{-4}$~s, while in the cosmological context 
$\bar\Omega^{-1}\sim 10^{10}$~years (this latter time scale might be 
considerably smaller since matter is {\it not} evenly distributed over the 
whole observable universe). 

We conclude with some final remarks. First, it is worth mentioning that in 
spite of the unbounded growth in Eqs.~(\ref{T00}-\ref{Tij}), 
$\left< T_{\mu \nu}\right>$ is {\it covariantly 
conserved}: $\nabla_\mu\left<T^\mu_\nu\right>=0$. In the static case 
[$f_{out}=f_{out}(\vec{x})$], for instance, this implies that the {\it total} 
vacuum energy is kept {\it constant}, although it continuously flows from 
spatial regions where its {\it density} is negative (and ever decreasing) to 
spatial regions where it is positive (and ever increasing). (This is an 
example of a spontaneous  time-like symmetry breaking.) Also, 
in the massless conformally coupled case ($m=0$ and $\xi=1/6$), 
the exponentially-increasing terms give {\it no} contribution to the 
anomalous value of the trace $\left<T^\mu_\mu\right>$. Finally, notice that
the exponential behavior appearing in Eqs.~(\ref{Phi2}-\ref{Tij}) leads only 
to {\it asymptotic} divergences; strictly speaking, all the quantities remain 
{\it finite} everywhere. This is in agreement, as it should be, with the 
seminal results obtained over the last decades on the topic of 
{\it renormalization} in QFTCS, which in summary show that a state 
(satisfying a positivity condition) which is {\it renormalizable} and free 
from {\it infrared divergences} at a particular time (i.e., with the only 
singular behavior of its two-point function being of a {\it Hadamard form}, 
for points in the same normal neighborhood of a given Cauchy surface), will 
remain so throughout the spacetime; {\it no divergences can appear due to a 
well-behaved evolution of the background 
spacetime}~\cite{Ford77,Fulling78,Radzi96}. This seminal result, whose 
importance cannot be stressed enough, seems to have discouraged further 
investigation on the topic of ``infrared behavior of fields in curved 
spacetime'' in the {\it general} context, as if it offered no more 
surprises. (For a thorough investigation in the case of de 
Sitter spacetime,
see Ref.~\cite{Anderson00}.) The vacuum-dominance effect 
presented here illustrates that this ``mathematical good behavior'' may still 
harbor interesting and wild physical phenomena. In fact, it is quite natural 
to expect that the infrared sector of a field theory should be very sensitive 
to the non-triviality of the background geometry, giving rise to legitimate 
QFTCS effects. We have made use of some idealizations (e.g., free scalar 
field, conformally-static asymptotic metrics) only to put in evidence the main 
idea 
behind the vacuum-dominance mechanism, avoiding unnecessary complications. 
The fact that this mechanism already manifests itself in such a simple and 
classically-well-behaved situation leads us to speculate that it might be of 
relevance in other, more complicated (and possibly realistic) 
scenarios (for instance, during the collapse of stars which classically would 
lead to the formation of black holes, or in the course of structure formation
during cosmological expansion).
Some of these legitimate QFTCS effects may still be waiting to be uncovered.

\acknowledgments

The authors would like to acknowledge partial financial support from 
Funda\c c\~ao de Amparo \`a Pesquisa do Estado de S\~ao Paulo (FAPESP) 
and Conselho Nacional de Desenvolvimento Cient\'\i fico e Tecnol\'ogico 
(CNPq). We thank George Matsas for valuable discussions and for reading 
the manuscript. D.V.\ would like to express special gratitude to Prof.\ 
Leonard Parker for extensive and illuminating discussions on QFTCS and its 
applications to cosmology, from where the author's interest on the 
``infrared regime'' of QFTCS has emerged.


\begin{thebibliography}{}

\bibitem{Parker68}
L.\ Parker, Phys.\ Rev.\ Lett.\ {\bf 21}, 562 (1968).

\bibitem{Parker69}
L.\ Parker, Phys.\ Rev.\ {\bf 183}, 1057 (1969).

\bibitem{BD82}
N.\ D.\ Birrell and P.\ C.\ W.\ Davies, {\it Quantum Fields in Curved Space}
(Cambridge Univ.\ Press, Cambridge, 1982).

\bibitem{Fulling89}
S.\ A.\ Fulling, {\it Aspects of Quantum Field Theory in Curved Spacetime}
(Cambridge Univ.\ Press, Cambridge, 1989). 

\bibitem{Wald94}
R.\ M.\ Wald, {\it Quantum Field Theory in Curved Spacetime and Black Hole
Thermodynamics} (Univ.\ of Chicago Press, Chicago, 1994).

\bibitem{Parker09}
L.\ Parker and D.\ Toms, {\it Quantum Field Theory in Curved Spacetime: Quantized
Fields and Gravity} (Cambridge Univ.\ Press, Cambridge, 2009).

\bibitem{Hawking74}
S.\ W.\ Hawking, Nature {\bf 248}, 30 (1974).

\bibitem{Hawking75}
S.\ W.\ Hawking, Commun.\ Math.\ Phys.\ {\bf 43}, 199 (1975).

\bibitem{PDG}
C.\ Amsler {\it et al.}, Phys.\ Lett.\ {\bf B 667}, 1 (2008).

\bibitem{Lattimer07}
J.\ M.\ Lattimer and M.\ Prakash, Phys.\ Rep.\ {\bf 442}, 109 (2007).

\bibitem{LMV}
W.\ C.\ C.\ Lima, G.\ E.\ A.\ Matsas, and D.\ A.\ T.\ Vanzella, in preparation.

\bibitem{Wald84}
R.\ M.\ Wald, {\it General Relativity} (Univ.\ of Chicago Press, Chicago, 1984).

\bibitem{Ford77}
L.\ H.\ Ford and L.\ Parker, Phys.\ Rev.\ D {\bf 16}, 245 (1977).

\bibitem{Fulling78}
S.\ A.\ Fulling, M.\ Sweeny, and R.\ M.\ Wald, Commun.\ Math.\ Phys.\ {\bf 63},
257 (1978).

\bibitem{Radzi96}
M.\ J.\ Radzikowski, Commun.\ Math.\ Phys.\ {\bf 180}, 1 (1996).

\bibitem{Anderson00}
P.\ R.\ Anderson, W.\ Eaker, S.\ Habib, C.\ Molina-Paris, and 
E.\ Mottola, Phys.\ Rev.\ D {\bf 62}, 124019 (2000).

\end{thebibliography}
\end{document}